# An oceanic basin oscillation-driving mechanism for tides

Yongfeng Yang (杨永峰)

Water Resources Comprehensive Development Center, Bureau of Water Resources of Shandong Province, Jinan, Shandong Province, China

Correspondence to: Yongfeng Yang (roufeng_yang@outlook.com)

**Abstract.** Tides represent the daily alternations of high and low waters along coastlines and in oceans, and the current theory (termed the gravitational forcing mechanism) explains them as a manifestation of the response of ocean water to the Moon's (Sun's) gravitational force. However, although the purely hydrodynamic models representing the current theory have been widely tested over global ocean, their tidal elevation accuracies are generally low. This implies an uncertainty as to whether the gravitational forcing mechanism is the best explanation for tides. In this study, we present a new theory (termed the oceanic basin oscillation-driving mechanism), in which tides are explained as a manifestation of oscillating ocean basin that is intricately linked to the elongated spinning solid Earth due to the Moon (Sun). Based on this new theory, we develop an algebraic tide model and test it using 11-year observations from 33 bottom pressure stations over the Pacific Ocean, the average Root Mean Square (*RMS*) deviation of tidal elevation predicted by this model against observation is 7.54 cm. Using a ratio of $M_2$ elevation *RMS* of ocean tide model EOT11a and its total tidal elevation *RMS* as a reference, we estimate the total tidal elevation *RMS* of six purely hydrodynamic models (i.e., Hallberg Isopycnal Model (HIM), OSU Tidal Inversion Software-GN (OTIS-GN), STORMTIDE model (STORMTIDE), OSU Tidal Inversion Software-ERB (OTIS-ERB), STM-1B, and HYbrid Coordinate Ocean Model (HYCOM)) to be 59.93, 51.64, 57.05, 38.56, 86.92, and 53.56 cm, respectively.

## 1. Introduction

From ancient times, the rhythmic rise and fall of water in coastal seas has attracted wide attention. Since these movements of water are closely related to frequent coastal activities, explaining the physics behind them has challenged human wisdom over the years. Aristotle (384-322 BC) attributed tides vaguely to the rocky nature of coastlines. Early Chinese people believed tides to be caused by the Earth's pulse or breathing, while others attributed tidal action to the different depths of ocean water. Galileo speculated that the rotation of the Earth around the Sun and its axis induced a motion in the sea to generate tides. Most of people, however, linked tidal action to the influences of the Moon and Sun. Seleucus (2nd century BC) was the first to propose this connection and thought that the tide's height correlated with the Moon's position relative to the Sun. A few Arabic theories proposed that the Moon heated and expanded the water using its rays. Descartes argued that tidal action formed when the Moon orbited the Earth, resulting in stresses between the ether and the Earth's surface. Kepler defined tidal action as the attraction of the Moon and Sun to the Earth's water. A relatively accurate solution to this issue was realized by Newton, who established that the Moon's gravitational attraction raises seawater to form tides. We here term the Newton's explanation as the attractive mechanism. This mechanism was subsequently developed through the work of these people such as Pierre-Simon Laplace, William Thomson, Baron Kelvin, Henri Poincaré, and Arthur Thomas Doodson, and finally reaches its present form. The attractive mechanism begins with a derivation of the tide-generating force, and this force is further decomposed into a tangential component and a vertical component. The tangential component, named the tractive force, exerts along the Earth's surface while the vertical component is balanced by the Earth's gravity. Based on these points, the present explanation for tides is by applying the tractive force to a watery Earth, taking into account the influences of these factors such as water depth, landmass, the Coriolis force, and friction. It assumes that the water's response to the force is slow and complicated, the upshot is that water moves like wave. A set of hydrodynamic equations of continuity and momentum are used to constrain the water movements around a rotating Earth. This is often referred as the dynamic tide theory, which is relative to the equilibrium tide theory. In this study, we term the dynamic tide theory as the gravitational forcing mechanism. Further elaboration on the dynamic tide and equilibrium tide theories is available in these works (Schureman, 1940; Doodson and Warburg, 1941; Cartwright, 1999; Deacon, 1971; Pugh and Woodworth, 2014; Gerkema, 2019).

Over the last five decades, the development of ocean tide models has facilitated global ground and spatial tide measurements, leading to the widespread study of both tidal dynamics and energy dissipation (Rochester, 1973; Miller, 1966; Munk, 1968; Riguzzi et al., 2010; Egbert and Ray, 2000; Egbert and Ray, 2003; Stammer et al., 2014). At the same time, internal tides have obtained more attention from scientific community (Gargett and Hughes, 1972; Phillips, 1974; Shepard, 1975; Feistel and Hagen, 1995; Roberts, 1975; Vlasenko et al., 2005; LeBlond and Mysak, 1978). Among these models, the data-constrained models (e.g., GOT, FES, EOT, DTU, TPXO) that are developed based on the harmonic analysis have realized a high accuracy of tidal elevation in the open ocean. In contrast, the purely hydrodynamic models such as HIM, OTIS-GN, STORMTIDE, OTIS-ERB, STM-1B, HYCOM, and Tidal Model forced by Ephemerides (TiME) that are developed based on the hydrodynamic equations are not accurate (Arbic et al., 2004; Green and Nycander, 2013; Müller et al., 2012; Egbert et al., 2004; Lyard et al., 2006; Chassignet et al., 2007; Weis et al., 2008; Hill et al., 2011). Many of these hydrodynamic models reported overtly that their $M_2$ tidal elevation accuracy is greater than 6.0 cm (Stammer et al., 2014). Conventionally, tide researchers treat a tidal elevation as a sum of the tidal elevations of many tidal constituents, and the $M_2$ tidal constituent is only one of them. Sulzbach et al. (2021) recently published an upgraded barotropic tidal model TiME, which was initially presented by Weis et al. (2008), the $M_2$ elevation *RMS* of this upgraded model is 3.39 cm. Such a high error level of individual tidal constituent indicates that the total tidal elevation accuracy of these hydrodynamic models may be very low. This present status implies an uncertainty as to whether the gravitational forcing mechanism is the best explanation for tides. The objects of this study is twofold, first, to present a new tidal theory, and, secondly, to take the new theory and the existing theory to see which of them is more suitable for accounting for the observed tides.

**2. Water movement in the oscillating container**

The movement of water in an oscillating container remains easily understood. As depicted in Figure 1(top), a declination of the water cup can lead the water level inside it to vary. Further, as depicted in Figure 1(middle and bottom), when the right end of a water box is elevated, the water flows towards the left. If we take line MN as a reference level, the water level at site M rises while the water level at site N drops. Conversely, when we restore the right end to its original position and elevate the left end, the water flows towards the right, causing the water level at site M to drop and the water level at site N to rise. By repeatedly elevating and dropping both ends, the water level fluctuations at sites M and N alternate. On the other hand, site S, located in the middle of the box, experiences minimal water level changes. From a perspective of energy transformation, elevating one end of the water box adds gravitational potential energy to the water. When the end falls, the potential energy is transformed into kinetic energy, causing the water to move. Beside these movements of water in oscillating containers, we also know that the earthquakes occurred in deep oceans have led ocean floor to shake, forming striking tsunamis. This argument of water movement indicates that an oscillating container may cause the water inside it to move back and forth, forming an all-around change in water level.

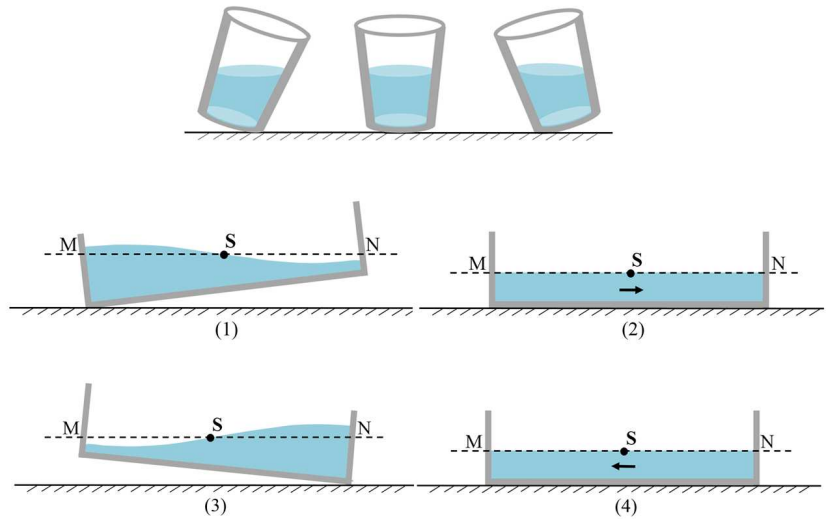

**Figure 1. Modeling the water movement in the oscillating container.** Top, the water cup is tilt orderly, causing water level inside it to vary. Middle and bottom, from (1), (2), (3) to (4), an orderly representation of the alternations of rise and fall of water in the oscillating water box. Arrows denote the directions of water movement.

## 3. An oceanic basin oscillation-driving mechanism for tides
### 3.1. Solid Earth deformation

The Earth simultaneously takes part in a multiple motion in the solar system: it spins around its axis, rotates around the barycenter of the Earth-Moon system, and revolves around the Sun. These movements are undoubtedly a consequence of different forces (i.e., gravitational and centrifugal forces). Accordingly, the Earth's body must deform in response to these external forces. It is well established that the Earth's spinning around its axis has caused its body to become a prolate sphere, in which the Earth's equator radius is greater than its polar radius. Besides this widely-known deformation, the gravitational force of the Moon and Sun has given the Earth's body another deformation, which is often called Earth tide or body tide. Since the time of Lord Kelvin, it has been believed that the yielding solid Earth due to the lunar (solar) gravitational force is an ellipsoid, in which solid Earth is slightly elongated along the Earth-Moon (Sun) line and shortened around the parts remote from the elongation (Love 1909).

Nevertheless, the ellipsoid's geometry (i.e., major semi-axis's length, minor semi-axis's length, and flattening) remains unresolved. Instead, the tidal displacement of a site is conventionally resolved through a combination of expanded potential equations and given Earth model.

Recently, Yang et al. (2024) simplified the Earth's dynamics by assuming the Earth to be a symmetrical, rotating, elastic, homogenous, and oceanless sphere. These authors proposed that the Earth's curved motions around the barycenter of the Earth-Moon system and around the Sun generate two centrifugal forces for the Earth. Using gravitational forces from the Moon and Sun to counterbalance these centrifugal forces, respectively, they concluded that the Earth's response to a combination of these opposing forces slightly elongates the Earth's body along the Earth-Moon (Sun) line and shortens it midway. The deformation could be represented geometrically with an ellipsoid whose major axis points to the Earth-Moon (Sun) line and whose center coincides with the Earth's center. By conceiving a rotating ellipse, these authors developed a geometric model in which both the ellipsoid's geometry and the tidal displacement of a site are resolved. According to these authors, the ellipsoids and their solution are illustrated with Figure 2, and the tidal displacement of a site at a time $t$ can be expressed as

$$H_{(t)} = ((R+M_e)^2 \cos^2\alpha + (R-M_s)^2 \sin^2\alpha)^{1/2} + ((R+S_e)^2 \cos^2\beta + (R-S_s)^2 \sin^2\beta)^{1/2} - 2R \quad (1)$$

Where $H_{(t)}$ is the vertical displacement of a site relative to the Earth's center in the ellipsoids due to the Moon and Sun at the time. $R$ is the mean radius of solid Earth, $M_e$ ($S_e$) and $M_s$ ($S_s$) are the elongation in the major semi-axis and the shortening in the minor semi-axis due to the Moon (Sun) at the time, respectively. $(R+M_e)$ and $(R-M_s)$ are the major semi-axis's length and minor semi-axis's length of the ellipsoid due to the Moon at the time, whereas $(R+S_e)$ and $(R-S_s)$ are the major semi-axis's length and minor semi-axis's length of the ellipsoid due the Sun at the time. $\alpha$ and $\beta$ are the lunar and solar angles of site P at the time. The lunar (solar) angle is the angle of a site and the Moon (Sun) relative to the Earth's center. More details of the geometric model and these parameters (i.e., $M_e$, $M_s$, $S_e$, $S_s$, $\alpha$, and $\beta$) may refer to Yang et al. (2024).

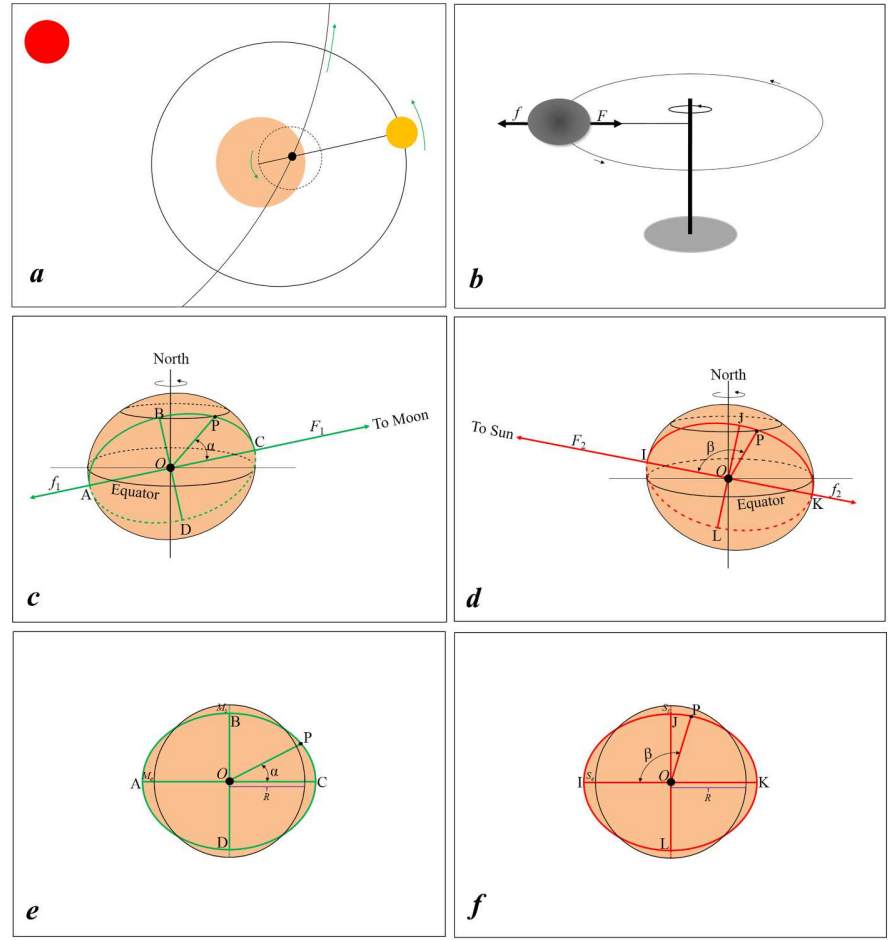

**Figure 2. Combined effect of opposing forces on solid Earth.** (*a*), the Earth's motions around the barycenter of the Earth-Moon system and around the Sun. (*b*), a sphere's deformation due to a combination of pull force and centrifugal force. (*c* and *d*), the ellipsoid due to a combination of the Moon's (Sun's) gravitational force and the centrifugal force. α and β are the lunar and solar angles of site P relative to the Earth's center *O*, respectively. (*e* and *f*), the sections dissected from the ellipsoid passing site P and the major axis AC (IK). Black circle denotes the original shape of section ABCD (IJKL). $M_e$ ($S_e$) and $M_s$ ($S_s$) denote the elongation and shortening in the ellipsoid due to the Moon (Sun), respectively. "Yang, Y., Zhang, Y., Liu, Q. et al, Scientific Reports, 14, 28527, 2024; licensed under a Creative Commons Attribution (CC BY)

license."

Yang et al. (2024) further showed that the averaged solid Earth elongation due to the Moon is about 2.7 times of the averaged solid Earth elongation due to the Sun, while the averaged solid Earth shortening due to the Moon is about 1.7 times of the averaged solid Earth shortening due to the Sun. These ratios imply that solid Earth is dominantly elongated along the Earth-Moon line and shortened around the parts remote from the elongation.

As shown in Figure 2 (c and d), an elongated solid Earth is represented spatially with two bulges that are along the Earth-Moon (Sun) line. These bulges necessitate a terrestrial location to move up and down as they are being entrained by the Earth's spinning. The movement of a location up or down corresponds to a fall or rise in gravity. Hence, changes in gravity can serve as an indicator to examine the existence of these bulges. In doing so, Yang et al. (2024) had constructed a correlation between lunar angle and gravity change for 20 superconducting gravimeter (SG) stations. A geographic distribution of the 20 SG stations are shown in Figure 3. The time covering of gravity data selected for each station is one month, and the data points are hourly. They found a decline in gravity during lunar angles of 0°-45° and 135°- 180°, where it represents that these stations fall into the double bugle mostly, and an increase in gravity during lunar phases of 45°-135°, where it represents these stations fall into the depression that is in the midway of the double bulge (Figure 4). These results confirm that solid Earth has been dominantly elongated along the Earth-Moon line.

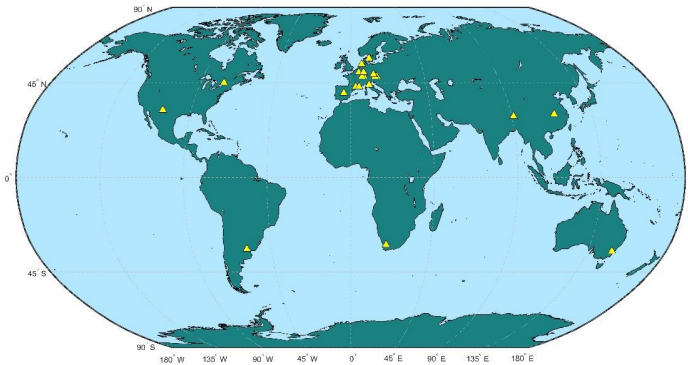

**Figure 3. A geographic distribution of the 20 SG stations (marked with yellow triangle).**

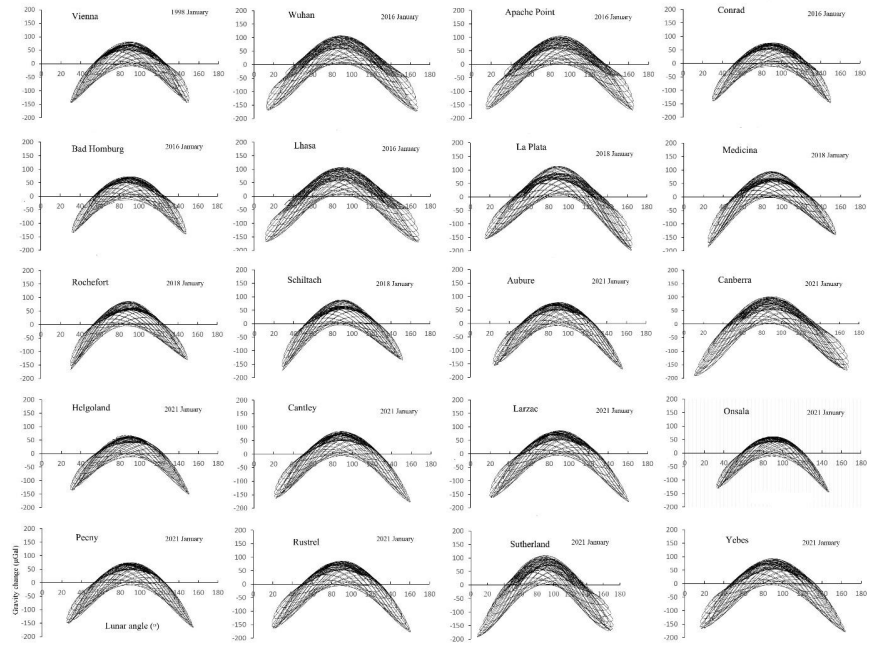

**Figure 4. Gravity change out to lunar angle.** Gravity data of 20 SG stations are selected from IGETs (Voigt et al., 2016). The lunar angle of a station is computed through the geographic latitudes and longitudes of the station and the Moon at the Earth's surface. The data points are hourly. "Yang, Y., Zhang, Y., Liu, Q. et al, Scientific Reports, 14, 28527, 2024; licensed under a Creative Commons Attribution (CC BY) license."

### 3.2. The elongated spinning solid Earth and its resultant tides

The Earth spins around its axis, this motion entrains the double bulge to regularly shake every part of ocean basin. As shown in Figure 5, the two solid bulges track from east to west, causing each part of ocean floor to move upwards and downwards. Here, we assume that ocean basins may be treated as naturally gigantic containers of water, and the water movements in the shaking ocean basins are similar to that in the oscillating container, which have been demonstrated in section 2.    Since the two solid bulges consecutively shake the ocean basin, water movement occurs regularly, the ocean surface thereby obtains two high waters and two low waters per day. Due to the existence of landmass, each part of the ocean surface is unnecessary to strictly follow two high waters and two low waters per day; in some of the extreme cases, there would occur one high water and one low water per day. In most time, the two solid bulges are located on the two sides of the equator, which leads to differences in the two shakes of the ocean basin, and further leads to unequal amplitudes for the two high or low waters per day. In particular, when the solid Earth deformation generated due to the Moon is associated with the solid Earth deformation generated due to the Sun, their combination would make water movement become strongest during full and new Moons, and become weakest during the first quarter and last quarter. This interaction generates two cycles of high and low waters over the course of a month (Fig. 6).

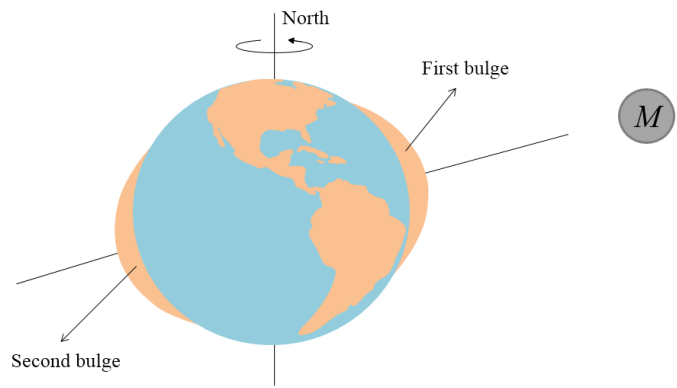

**Figure 5. Conceptual model of the elongated spinning solid Earth.** Oceans are marked with shallow blue, while continents or the double bulge are marked with yellow.

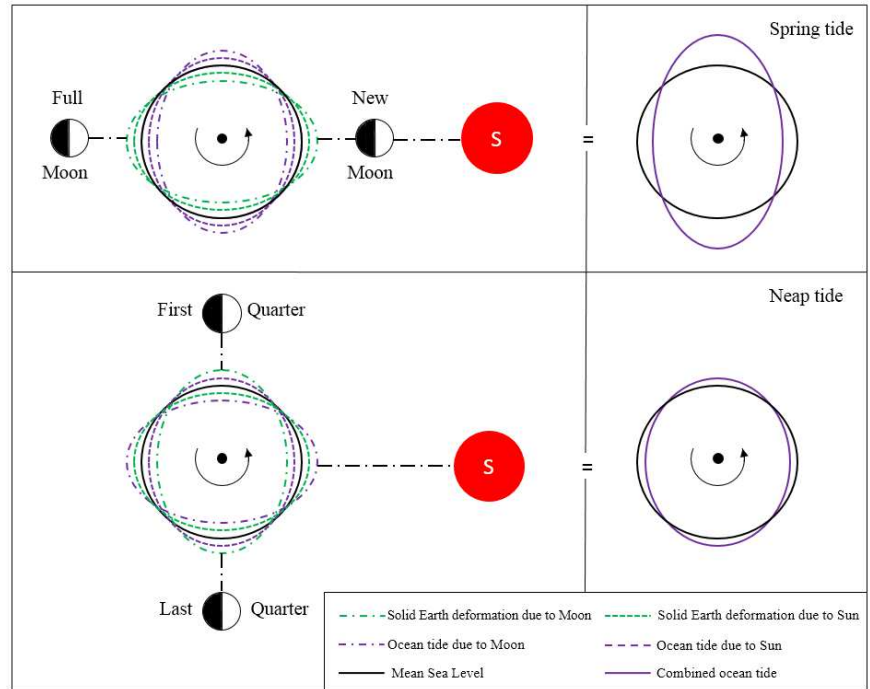

**Figure 6. Conceptual model of Spring-Neap tides under the relative motions of the Moon and Sun.** Top, the Moon, the Sun, and the Earth are in line at the moment. Bottom, the Moon is at the first quarter and at the last quarter, respectively.

### 3.3. Oceanic Basin Oscillation Tide Model (OBOTM)

The above demonstration illustrates that the elongated spinning solid Earth may cause ocean water to move back and forth. However, the detail of how water begins to move in basin deserves further discussion. See Figure 7 (a), when a site in the ocean floor rises, the water above it instantly rises too, while the water above the falling site in the ocean floor immediately drops. This variably affects the gravitational potential of the ocean surface everywhere. For instance, site A experiences a decrease in potential, whereas site B experiences an increase in potential. Ocean water is a continuous mass, and its potential corresponds to a pressure. Here, we assume that under the effect of this pressure difference, water initiates a horizontal movement.

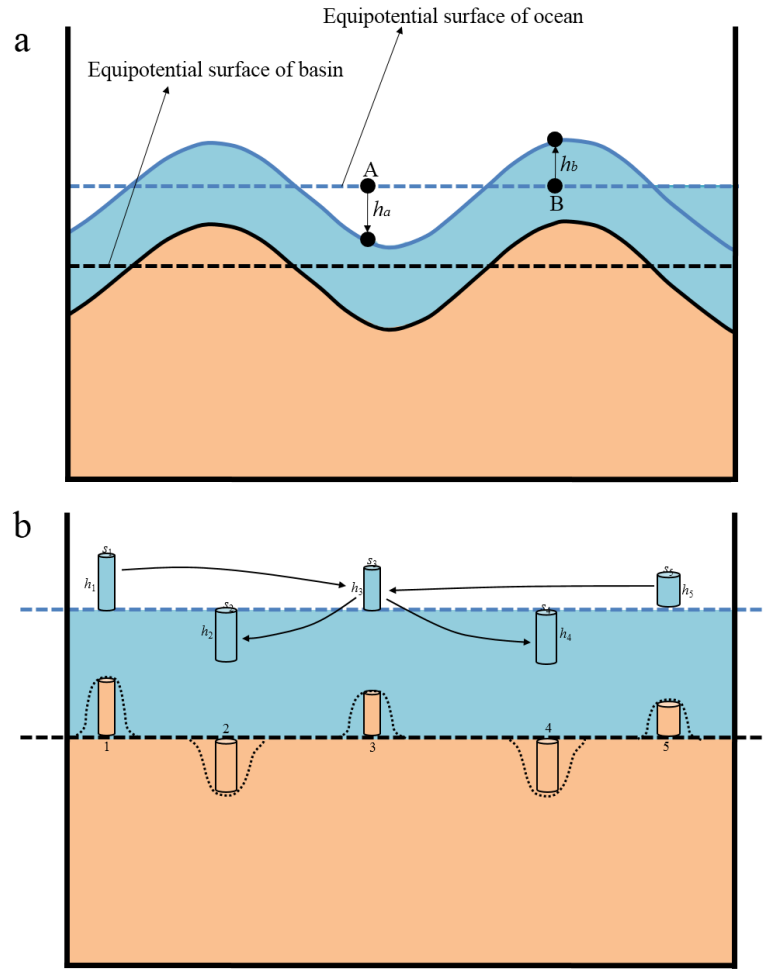

**Figure 7. (a) Conceptual model of instant adjustments of ocean surface due to ocean floor deformation.** $h_a$ ($h_b$) denotes the instant drop (rise) in water level at site A(B). Note that the deformations of ocean surface and floor are highly exaggerated. **(b) Conceptual model of water transferring between sites**. $h_1$, $h_2$, $h_3$, $h_4$, and $h_5$ denote the heights of resultant water depth changes due to the instant drop (rise) of the sites in the ocean floor. $s_1$, $s_2$, $s_3$, $s_4$, and $s_5$ denote the base areas of water depth changes. Lines with arrow denote the water travelling directions between the region above site 3 and other regions above sites 1, 2, 4, and 5. Dashed line regions represent the rises and drops of the ocean floor at the responding sites.

The water movements within ocean basin mean that changes in water depth at any given site are determined by changes in water depth at all other sites and at the site itself. See Figure 7 (b), we assume that the rises (drops) of the sites in the ocean floor give rise to instant water depth changes in the ocean. We further assume that the water level changes above sites 1 and 5 lead water to move towards site 3, and the water level changes above sites 3 and 2(4) lead water to move towards site 2(4). Of course, the tidal force (if it is true) exerts effect on water, and these factors (i.e., ocean depth, landmass, the Coriolis force, and friction) have influences on the travelling water. We here neglect the effect of tidal force and these factors, and consider the time of water travelling between two sites, then, the final water depth change at site 3 may be simply written as $\Delta h_{(t0)} = h_{1(t1-t0)}s_1/s_3 + h_{2(t2-t0)}s_2/s_3 + h_{4(t4-t0)}s_4/s_3 + h_{5(t5-t0)}s_5/s_3 - h_{3(t0)}$, where $h_{1(t1-t0)}$, $h_{2(t2-t0)}$, $h_{4(t4-t0)}$, and $h_{5(t5-t0)}$ denote the water depth changes above the sites 1, 2, 4, and 5 at the corresponding time ($t1$-$t0$), ($t2$-$t0$), ($t4$-$t0$), and ($t5$-$t0$), respectively, $h_{3(t0)}$ denotes the water depth change above site 3 at the time ($t0$). For example, if the UTC time at site 3 is 6:00 of today, and the duration that the travelling water takes from site 1 to reach site 3 is 8 hours, then, the $h_{1(t1-t0)}$ denotes the water depth change above site 1 at the time 22:00 of yesterday. $s_1$, $s_2$, $s_3$, $s_4$, and $s_5$ denote the base areas of water depth changes, and thereby $h_1s_1$, $h_2s_2$, $h_2s_2$, $h_3h_3$, $h_4s_4$, and $h_5s_5$ denote the volumes of corresponding water columns, respectively. We now assume that the whole ocean floor may be divided into countless sites, and a rise or drop of each of these sites may give rise to an instant water depth change, and that the water travelling between any two sites takes some time, then the final water depth change at any given site at time ($t0$) can be written as

$$\Delta h_{(t0)} = \sum_{i=1}^{n} h_{i(ti-t_0)} s_i / s - h_{(t0)} \qquad (2)$$

where $h_{(t0)}$ is the water depth change above the given site at the time ($t0$), $h_{i(ti-t0)}$ is the water depth change above the $i$th at the time ($ti$-$t0$). Using the latitudes and longitudes of these sites, we may work out a geographical distance $L$ between any two sites. If we treat water movement in the ocean as tidal wave propagation, then, the duration may be written as $t=L/v$, where $v$ is the tidal wave speed. In this study, we adapt $v$=715 km per hour (Pugh and Woodworth, 2014). Although water movement may be deflected by the Coriolis force, its actual path is hard to accurately track. We here use a geographical distance of two sites to approximately represent the length of the actual path of the travelling water. We further term $s_i/s$ as $k_n$, then, the equation (2) can be converted into

$$\Delta h_{(t0)} = \sum_{i=1}^{n} k_i h_{i(ti-t_0)} - h_{(t0)} \qquad (3)$$

Equation (3) indicates that, if we assume the water depth change above a site to be equal to the vertical displacement of the site in the ocean floor, and if the water depth change above a given site in the ocean floor can be measured, we may use the vertical displacement data of these sites in the ocean floor and the measured water depth change data above the given site as input, and take a least-squares fitting to resolve these parameters $k_1$, $k_2$, $k_3$, …, and $k_n$. With the known parameters and the known vertical displacement of related sites in the ocean floor in the future (past) time, we may extrapolate the water depth change above the given site in the future (past) time. This relationship provides a theoretical foundation for tide prediction.

About 29% of the Earth's surface is covered with landmass, the obstacle of landmass inevitably limits water travelling between ocean basins. Consequently, we cannot guarantee a global applicability for equation (3). Alternatively, a local test may be promising. Here, we select the Pacific Ocean as the experimental region and divide it into a 10º resolution (Fig. 8). To simplify following calculations, we use the locations of nodes of the grid, which is between 30º N and 30º S, to act as the primary sites in the ocean floor that can produce the water depth changes, and a total of 92 sites are extracted from this grid. The latitude and longitude of these 92 sites are listed in Table 1. This choice of latitudinal belt is based on the fact that the Moon's position mostly transfers between 18º N and 18º S. The grid is plotted with reference to both the equator and meridian, which allows the latitude and longitude of these sites to be easily known. And then, the water depth change above any given site in the Pacific Ocean floor may be written as

$$\Delta h_{(t0)} = \sum_{i=1}^{92} k_i h_{i(ti-t_0)} - h_{(t0)} \qquad (4)$$

Where $h_{(t0)}$ represent the vertical displacement of the given site in the Pacific Ocean floor at the time ($t0$), $h_{i(ti-t0)}$ represent the vertical displacements of the $i$th site in the Pacific Ocean floor at the time ($ti$-$t0$).

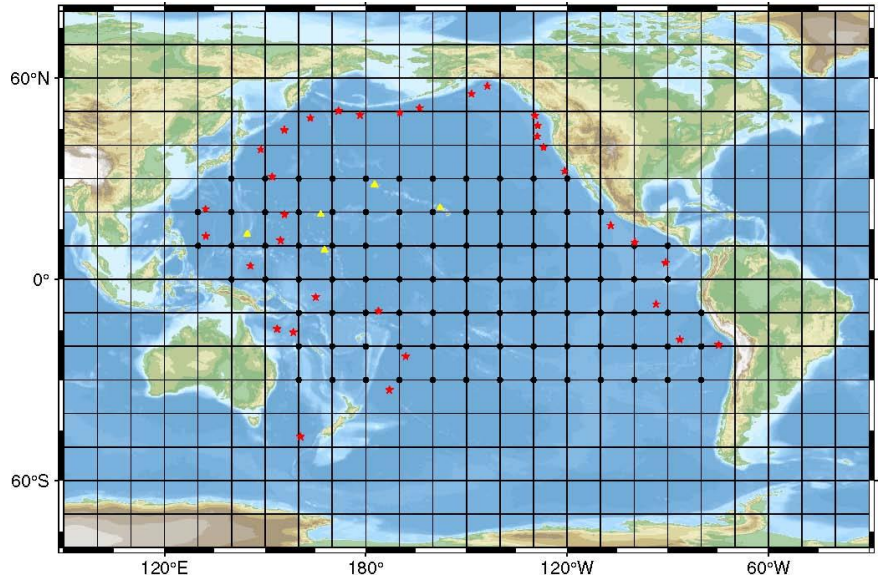

**Figure 8. Geographical distribution of 92 primary sites and 38 given sites over the Pacific Ocean.** The primary sites (marked with black dot) are located at the nodes of the grid. The 38 given sites include 33 DART bottom pressure stations (marked with red star) and 5 tide-gauge stations (marked with yellow triangle).

**Table 1. Geographical information of 92 primary sites in the Pacific Ocean floor**

| No. | Latitude(°) | Longitude(°) | No. | Latitude(°) | Longitude(°) | No. | Latitude(°) | Longitude(°) |
|---|---|---|---|---|---|---|---|---|
| 1 | 10 | 130 | 32 | -30 | 190 | 63 | 0 | 230 |
| 2 | 20 | 130 | 33 | -20 | 190 | 64 | 10 | 230 |
| 3 | 0 | 140 | 34 | -10 | 190 | 65 | 20 | 230 |
| 4 | 10 | 140 | 35 | 0 | 190 | 66 | 30 | 230 |
| 5 | 20 | 140 | 36 | 10 | 190 | 67 | -30 | 240 |
| 6 | 30 | 140 | 37 | 20 | 190 | 68 | -20 | 240 |
| 7 | 0 | 150 | 38 | 30 | 190 | 69 | -10 | 240 |
| 8 | 10 | 150 | 39 | -30 | 200 | 70 | 0 | 240 |
| 9 | 20 | 150 | 40 | -20 | 200 | 71 | 10 | 240 |
| 10 | 30 | 150 | 41 | -10 | 200 | 72 | 20 | 240 |
| 11 | -30 | 160 | 42 | 0 | 200 | 73 | 30 | 240 |
| 12 | -20 | 160 | 43 | 10 | 200 | 74 | -30 | 250 |
| 13 | -10 | 160 | 44 | 20 | 200 | 75 | -20 | 250 |
| 14 | 0 | 160 | 45 | 30 | 200 | 76 | -10 | 250 |
| 15 | 10 | 160 | 46 | -30 | 210 | 77 | 0 | 250 |
| 16 | 20 | 160 | 47 | -20 | 210 | 78 | 10 | 250 |
| 17 | 30 | 160 | 48 | -10 | 210 | 79 | 20 | 250 |
| 18 | -30 | 170 | 49 | 0 | 210 | 80 | -30 | 260 |
| 19 | -20 | 170 | 50 | 10 | 210 | 81 | -20 | 260 |
| 20 | -10 | 170 | 51 | 20 | 210 | 82 | -10 | 260 |
| 21 | 0 | 170 | 52 | 30 | 210 | 83 | 0 | 260 |
| 22 | 10 | 170 | 53 | -30 | 220 | 84 | 10 | 260 |
| 23 | 20 | 170 | 54 | -20 | 220 | 85 | -30 | 270 |
| 24 | 30 | 170 | 55 | -10 | 220 | 86 | -20 | 270 |
| 25 | -30 | 180 | 56 | 0 | 220 | 87 | -10 | 270 |
| 26 | -20 | 180 | 57 | 10 | 220 | 88 | 0 | 270 |
| 27 | -10 | 180 | 58 | 20 | 220 | 89 | 10 | 270 |
| 28 | 0 | 180 | 59 | 30 | 220 | 90 | -30 | 280 |
| 29 | 10 | 180 | 60 | -30 | 230 | 91 | -20 | 280 |
| 30 | 20 | 180 | 61 | -20 | 230 | 92 | -10 | 280 |
| 31 | 30 | 180 | 62 | -10 | 230 | | | |

Water depth changes in oceans have been extensively measured by bottom pressure recorder and satellite altimetry, and the formed data are freely accessible at some archives such as NDBC (National Data Buoy Center) (https://www.ndbc.noaa.gov/) and AVISO (Archiving, Validation and Interpretation of Satellite Oceanographic) (https://www.aviso.altimetry.fr/). As mentioned earlier, the tidal displacement of a site is conventionally computed through a solid tide model that incorporates the expanded potential equations and given Earth model. The presently widely-used solid tide model is that recommended by the IERS (International Earth Rotation System) conventions (2010) (Gérard and Luzum, 2010). Notwithstanding, Yang et al. (2024)

have revealed that the solid tide model recommended by the IERS conventions (2010) has a vertical displacement error of greater than 9.0 cm, and suggested that such a large error has imposed a challenge on its application in various fields. Alternatively, the geometric solid tide model presented by these authors has a vertical displacement error of close to 2.0 cm. This excellence allows us to choose the newly proposed geometric model to compute the vertical displacement of any site in the Pacific Ocean floor. Such computation is proceeded through equation (1), which is already exhibited in section 3.1.

To help readers to easily produce tidal displacement at any site, we have developed equation (1) into an excel code that is included in the supplementary material of this research. Two set of observations are used to test equation (4) separately, and the results are evaluated in terms of the root mean square (*RMS*) of water depth change (i.e., tidal height). The *RMS* deviation for a gauge site may be written as follows:

$$RMS = \sqrt{\sum_{i=1}^{m}\left(\Delta Y_{(t)-predicted} - \Delta Y_{(t)-observed}\right)^2 / m} \qquad (5)$$

where $\Delta Y_{(t)-predicted}$ and $\Delta Y_{(t)-observed}$ represent the predicted and observed water depth changes (i.e., tidal heights), respectively, and *m* is the total number of data points used in the modelling.

3.3.1 Bottom pressure data

Bottom pressure data provide direct information on the water depth changes over gauge stations and are being widely used in earth science field. Ray (2013) made a comparison of bottom-pressure and altimetric ocean tides, and totally 151 gauge stations were employed. The data of these stations were subsequently used to evaluate various ocean tide models (Stammer et al., 2014). In the supplementary file of Ray (2013), the constants (i.e., amplitude and phase) of tidal constituents over these stations were established, but these constituents cannot be directly used to generate real-time data. We note that the bottom pressure stations employed by Ray (2013) come from several sources such as IAPSO, GLOUP, MOVE, ASTTEX, HOME, RAPID, DART, and so on, and DART is the primary source that has included 47 stations. However, the observations of most of these 151 stations cannot be presently/easily found. In contrast, the observations of DART stations are publicly available. Only 36 of these 47 DART stations belong to the Pacific Ocean region, but 3 of these 36 DART stations, which are DART_d125, DART_d157, and DART_d171, are presently not included in NDBC. Finally, the observations of 33 DART stations are selected for this study. It is worth noting that the quality of DART data is not too good, and their usage requires some attention. For example, the suspect and missing data frequently appear in the time series of most of these stations. More details of DART data and the treatment on them are carefully introduced in Ray (2013). Also note, the pressure data in NDBC are recorded with 15 min span. In this study, we only extract hourly data for analysis. The usage of hourly data means that we have to carefully choose the water depth change above the sites at the corresponding time. In the expression of the duration $t=L/v$, $v$=750 km/h, but the geographic distance of two sites is computed through their latitudes and longitudes, consequently, the amount of the duration $t$ needs to be rounded. For instance, if $t= L/v$ =3.4 or 3.1, we round it to 3.0, and if $t= L/v$ =3.5 or 3.9, we round it to 4.0. The information of 33 DART stations and their time series selected are listed in Table 2. A geographical distribution of these stations is exhibited in Figure 8. First, we use hourly water depth change data of two months and hourly vertical displacement data of same duration to resolve the parameters of equation (4). The water depth change is computed through subtracting mean water depth from the observed water depth, the vertical displacement is computed through equation (1). Then, we use equation (4) of known parameters and vertical displacement data to reproduce the water depth changes in the future time and compared them to observations. The hourly series used to resolve equation (4) cover the years of 2007-2010. The prediction series cover the years of 2007-2010 (all 33 stations are involved) and the year of 2021 (only 25 stations are involved, because other 8 stations during this period are either without data or seriously low quality of data). Table 2 exhibits a comparison of prediction and observation, with some comparisons shown in Figure 9. Over these 33 stations that cover the years of 2007-2010, their averaged *RMS* of prediction against observation is 6.9 cm, while over these 25 stations that cover the year of 2021, their averaged *RMS* is 7.54 cm. The results of two prediction series are close, which proves the model to be constantly effective.

**Table 2. OBOTM tide elevations against bottom pressure data in the Pacific Ocean**

| Bottom pressure station | | | Observed series for resolving the parameters of equation (4) | Predicted series | RMS (cm) |
| --- | --- | --- | --- | --- | --- |
| ID | Latitude (º) | Longitude (º) | | | |
| DART_21413 | 30.55 | 152.12 | 2008.8-2008.9 | 2008.10-2008.12 | 7.25 |
| DART_21414 | 48.94 | 178.28 | 2008.1-2008.2 | 2009.7-2009.8 (2021.1-2021.10) | 4.42 (5.15) |
| DART_21415 | 50.18 | 171.85 | 2008.4-2008.5 | 2008.6-2008.8 (2021.3-2021.7) | 4.84 (4.84) |
| DART_21416 | 48.04 | 163.49 | 2008.4-2008.5 | 2008.6-2008.8 (2021.1-2021.12) | 7.14 (7.41) |
| DART_21418 | 38.71 | 148.69 | 2010.1-2010.2 | 2010.3-2010.8 (2021.7-2021.12) | 7.12 (10.14) |
| DART_21419 | 44.46 | 155.74 | 2010.7-2010.8 | 2010.9-2010.12 (2021.2-2021.7) | 5.08 (5.76) |
| DART_32401 | -19.55 | 285.19 | 2008.1-2008.2 | 2008.3-2008.9 (2021.1-2021.8) | 4.14 (4.65) |
| DART_32411 | 4.92 | 269.32 | 2008.10-2008.11 | 2008.12-2009.4 (2021.1-2021.12) | 7.20 (7.45) |
| DART_32412 | -17.98 | 273.61 | 2008.1-2008.2 | 2008.3-2008.12 | 3.37 |
| DART_32413 | -7.40 | 266.50 | 2008.1-2008.2 | 2008.3-2008.11 (2021.1-2021.12) | 4.31 (6.17) |
| DART_43412 | 16.03 | 253.00 | 2010.6-2010.7 | 2010.9-2010.12 (2021.1-2021.8) | 6.00 (5.40) |
| DART_43413 | 11.07 | 260.15 | 2008.1-2008.2 | 2008.3-2008.11 | 4.31 |
| DART_46402 | 51.07 | 195.99 | 2008.6-2008.7 | 2008.3-2008.11 (2021.1-2021.12) | 6.21 (9.26) |
| DART_46404 | 45.86 | 231.22 | 2008.2-2008.3 | 2008.4-2008.12 (2021.1-2021.12) | 11.17 (17.83) |
| DART_46407 | 42.60 | 231.10 | 2007.6-2007.7 | 2007.8-2007.11 (2021.1-2021.12) | 7.45 (7.80) |
| DART_46408 | 49.63 | 190.13 | 2008.6-2008.7 | 2008.9-2008.12 (2021.7-2021.12) | 6.32 (6.82) |
| DART_46409 | 55.30 | 211.49 | 2008.6-2008.7 | 2008.9-2008.11 (2021.7-2021.12) | 10.24 (9.22) |
| DART_46410 | 57.64 | 216.21 | 2007.2-2007.3 | 2007.4-2007.12 (2021.3-2021.5) | 11.86 (11.02) |
| DART_46411 | 39.33 | 232.99 | 2007.1-2007.2 | 2010.10-2010.12 (2021.1-2021.3) | 8.59 (8.37) |
| DART_46412 | 32.25 | 239.30 | 2008.4-2008.5 | 2008.6-2008.12 | 5.40 |
| DART_46419 | 48.76 | 230.38 | 2009.6-2009.7 | 2010.6-2010.9 (2021.1-2021.5) | 10.00 (9.94) |
| DART_51425 | -9.49 | 183.76 | 2008.3-2008.4 | 2008.5-2008.12 (2021.1-2021.12) | 12.90 (11.98) |
| DART_51426 | -22.99 | 191.90 | 2008.3-2008.4 | 2008.5-2008.12 | 8.30 |
| DART_52401 | 19.29 | 155.77 | 2008.6-2008.7 | 2008.8-2008.12 (2021.1-2021.12) | 7.90 (5.54) |
| DART_52402 | 11.57 | 154.58 | 2008.6-2008.7 | 2008.8-2008.12 | 3.56 |
| DART_52403 | 4.03 | 145.60 | 2009.6-2009.7 | 2010.5-2010.8 (2021.8-2021.10) | 3.80 (3.53) |
| DART_52404 | 20.95 | 132.22 | 2008.6-2008.7 | 2008.8-2008.12 (2021.1-2021.6) | 7.03 (5.41) |
| DART_52405 | 12.88 | 132.33 | 2008.6-2008.7 | 2008.8-2008.12 | 10.38 |
| DART_52406 | -5.33 | 165.08 | 2008.6-2008.7 | 2008.8-2008.12 (2021.1-2021.12) | 9.09 (8.17) |
| DART_54401 | -33.01 | 187.02 | 2010.1-2010.2 | 2010.3-2010.7 | 5.87 |
| DART_55012 | -15.80 | 158.40 | 2008.8-2008.9 | 2008.10-2008.12 (2021.1-2021.9) | 3.52 (3.86) |
| DART_55015 | -46.92 | 160.56 | 2009.4-2009.5 | 2009.6-2009.12 (2021.1-2021.12) | 5.37 (5.52) |
| DART_55023 | -14.80 | 153.58 | 2010.10-2010.11 | 2010.12 (2021.1-2021.12) | 7.52 (7.16) |
| | | | | Mean | 6.90 (7.54) |

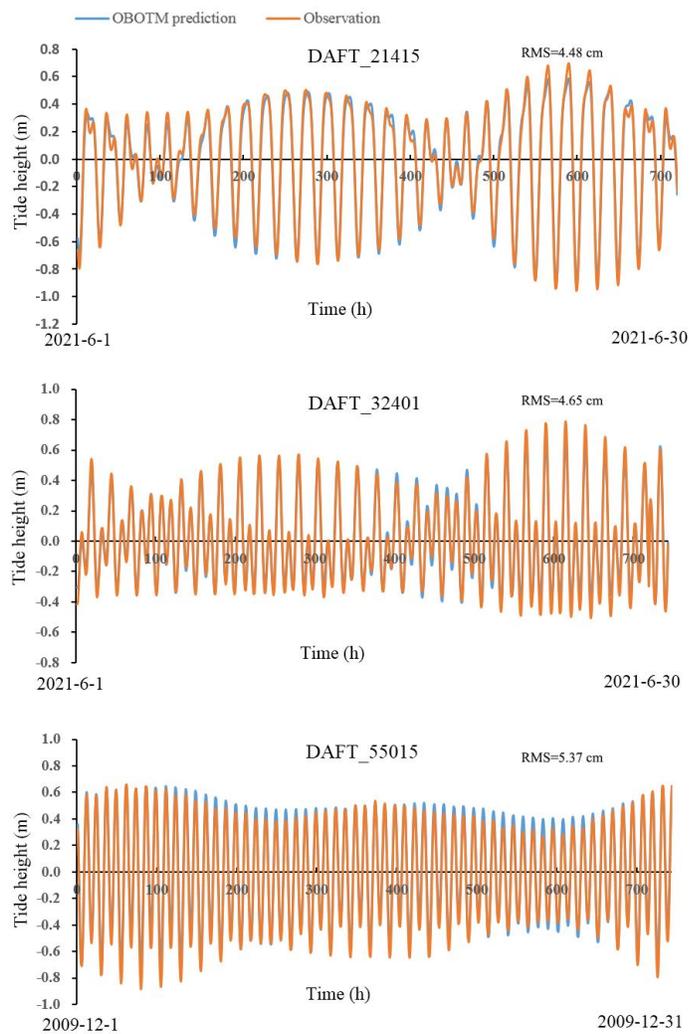

**Figure 9. Water depth change (tidal elevation) comparison of prediction and observation.** The predictions are made by OBOTM, while the observations are from DART bottom pressure stations. The data points are hourly.

To examine the strength of this model, we compare it with the purely hydrodynamic models. As introduced in these works (e.g., Pugh, 1987; Fok, 2012; Gerkema, 2019), the hydrodynamic models are developed based on a set of equations of continuity and momentum as below:

$$\begin{cases} \partial\zeta/\partial t + (\partial/\partial x)(Du) + (\partial/\partial y)(Dv) = 0 \\ \partial u/\partial t + v(\partial u/\partial y) + u(\partial u/\partial x) - fv = -(\partial\Omega/\partial x) - (1/\rho)(\partial P/\partial x - \partial F/\partial z) \\ \partial v/\partial t + u(\partial v/\partial x) + v(\partial v/\partial y) + fu = -(\partial\Omega/\partial y) - (1/\rho)(\partial P/\partial y - \partial G/\partial z) \end{cases} \quad (6)$$

where $\zeta$ and $\rho$ are sea level elevation (i.e., tidal height) and water density, respectively. More details of these equations may refer to that work (Pugh, 1987). By assuming boundary conditions, the goal of the hydrodynamic models is to resolve the sea level elevation $\zeta$ and flow velocities $u$ and $v$. This method may date back to the time of Laplace. Although it has passed almost 250 years, a precise solution on these equations is still not yet realized. In most of cases, the hydrodynamic models mostly reported the elevation of $M_2$ constituent (see Stammer et al. (2014)). But what's the meaning of $M_2$ elevation *RMS* in the total tidal elevation? Usually, tide researchers treat the observed tidal elevation as a sum of the elevations of many tidal constituents. For the $M_2$ tidal constituent, for instance, if the observed tidal elevation of this constituent is 0.3 m, and if a tide model produces a $M_2$ elevation of 0.26 m, then, the $M_2$ elevation *RMS* of production against observation for this model is 0.04 m. From these points, we understand that the $M_2$ elevation *RMS* of hydrodynamic model cannot provide a precise information about its total tidal elevation *RMS*. This leaves the total tidal elevation accuracy of the model to be unclear. Even so, we still may find a way to estimate it. Stammer et al. (2014) evaluated three types of ocean tide models: empirical, purely hydrodynamic, and data-constrained. Unlike purely hydrodynamic models that mostly discuss the $M_2$ constituent, the 8 major constituents are usually involved in these data-constrained models. We also note that, except for the elevation *RMS*s of the 8 constituents and their *RSS* (root-sum-square), the total tidal elevation *RMS*s of these data-constrained models (FES2014 and EOT20, for instance) are less overtly reported (see Lyard et al. (2021), Hart-Davis et al. (2021), and Stammer et al. (2014)). We feel that it is necessary to test the total tidal elevation accuracies of data-constrained models. In doing so, the tidal elevations at the locations of 20 bottom pressure stations over the year of 2021 are firstly produced through three data-constrained models (i.e., EOT11a, EOT20, and Fes2014b), and then are compared with the observed tidal elevations. The information of these stations and their comparisons are listed in Table 3. These stations also belong to the Pacific Ocean region and are already included in our modelling (see Table 1). The time series of these stations, which were already used in Ray (2013) and Stammer et al. (2014), are from 2007 to 2010. Unlike the coastal tides that vary over time, the ocean tides are relatively constant, this means that the elevation accuracies of these models over the years of 2007-2010 may be equal to that over the year of 2021. We find that the averaged total tidal elevation *RMS* of reproduced data against observed data over these 20 stations for EOT11a, EOT20, and Fes2014b are 3.88, 4.19, and 3.97 cm, respectively. The elevation accuracies of three models are nearly close, even though the publication of EOT11a is earlier about 10 years than that of EOT20 and Fes2014b. As the observed tidal elevation is conventionally treated as a sum of the elevations of many tidal constituents, the ratio of $M_2$ elevation *RMS* and total tidal elevation *RMS* of a model may be treated as a reference to estimate the total tidal elevation *RMS* of another mode, so long as the $M_2$ elevation *RMS* of another model is known. Here, we use the ratio of $M_2$ elevation *RMS* and total tidal elevation *RMS* of EOT11a to estimate the total tidal elevation *RMS* of 6 purely hydrodynamic models (HIM, OTIS-GN, STORMTIDE, OTIS-ERB, STM-1B, and HYCOM), which were involved in Stammer et al. (2014). See Table 3, the ratio of $M_2$ elevation *RMS* and total tidal elevation *RMS* of EOT11a is 0.56/3.88=0.146. The $M_2$ elevation *RMS*s of these hydrodynamic models and their total tidal elevation *RMS*s estimated are listed in Table 4. For HIM, OTIS-GN, STORMTIDE, OTIS-ERB, STM-1B, and HYCOM, their total tidal elevation *RMS*s estimated are 59.93, 51.64, 57.05, 38.56, 86.92, and 53.56 cm, respectively. The averaged tidal elevation *RMS* of these models is 57.95 cm. Sulzbach et al. (2021) recently reported a $M_2$ elevation *RMS* of 3.39 cm for an upgraded barotropic hydrodynamic tidal model TiME, which was initially presented by Weis et al. (2008). Using the ratio of EOT11a, we work out the total tidal elevation *RMS* of the upgraded TiME to be 23.22 cm. As seen in Stammer et al. (2014), the *RSS*s of two data-constrained models (i.e., EOT11a and FES12) against bottom pressure recorders are respectively 1.056 and 1.120 cm. However, as listed in Table 3, the total tidal elevation *RMS*s of three data-constrained models (i.e., EOT11a, EOT20, and Fes2014b) are close to 4.0 cm. Please note, Fes2014b is also an upgraded version of FES12. Hence, the total tidal elevation *RMS* of a model may be many times of the *RSS* of 8 major tidal constituents of the model, for EOT11a, the ratio of *RMS* and *RSS* is approximately 4.0. The hydrodynamic model HIM (Arbic et al., 2004) had reported its *RSS* of 8 major constituents, which is 11.14 (10.36) cm. If we use this ratio of *RMS* and *RSS* of EOT11a as a reference, the total tidal elevation *RMS* for HIM may reach 44.56 (41.44) cm. On the whole, the tidal elevation accuracy of purely hydrodynamic models is rather low, because most of the bottom pressure stations selected in this study have an observed tidal elevation of less than 1.0 m.

**Table 3. Data-constrained ocean tide models' elevations against bottom pressure data in the Pacific Ocean**

| Bottom pressure station | | | Reproduction series | Total tide elevation RMS (cm) | | | $M_2$ tide elevation RMS (cm) |
|---|---|---|---|---|---|---|---|
| ID | Latitude (°) | Longitude (°) | | EOT11a | EOT20 | Fes2014b | EOT11a |
| DART_21414 | 48.94 | 178.28 | 2021.4-2021.10 | 4.11 | 3.63 | 4.08 | |
| DART_21415 | 50.18 | 171.85 | 2021.3-2021.10 | 3.02 | 4.07 | 3.10 | |
| DART_21416 | 48.04 | 163.49 | 2021.1-2021.12 | 3.70 | 4.58 | 3.70 | |
| DART_21419 | 44.46 | 155.74 | 2021.1-2021.7 | 3.71 | 4.10 | 3.67 | |
| DART_32401 | -19.55 | 285.19 | 2021.1-2021.12 | 2.65 | 2.93 | 2.70 | |
| DART_32411 | 4.92 | 269.32 | 2021.1-2021.12 | 3.13 | 3.23 | 2.95 | |
| DART_32413 | -7.40 | 266.50 | 2021.1-2021.5 | 1.96 | 2.42 | 2.00 | |
| DART_43412 | 16.03 | 253.00 | 2021.1-2021.12 | 2.98 | 3.04 | 2.74 | |
| DART_46402 | 51.07 | 195.99 | 2021.1-2021.7 | 8.07 | 8.44 | 8.03 | |
| DART_46404 | 45.86 | 231.22 | 2021.1-2021.12 | 3.12 | 4.14 | 6.00 | |
| DART_46407 | 42.60 | 231.10 | 2021.1-2021.12 | 4.50 | 3.27 | 4.44 | |
| DART_46410 | 57.64 | 216.21 | 2021.3-2021.5 | 3.71 | 3.70 | 3.80 | |
| DART_51425 | -9.49 | 183.76 | 2021.1-2021.12 | 7.35 | 8.17 | 7.27 | |
| DART_52401 | 19.29 | 155.77 | 2021.1-2021.7 | 2.32 | 2.63 | 2.35 | |
| DART_52402 | 11.57 | 154.58 | 2021.1-2021.8 | 2.12 | 3.23 | 2.22 | |
| DART_52405 | 12.88 | 132.33 | 2021.8-2021.12 | 6.57 | 4.95 | 6.53 | |
| DART_52406 | -5.33 | 165.08 | 2021.1-2021.12 | 4.28 | 5.43 | 4.16 | |
| DART_55012 | -15.80 | 158.40 | 2021.1-2021.9 | 3.54 | 3.78 | 3.07 | |
| DART_55015 | -46.92 | 160.56 | 2021.1-2021.12 | 3.21 | 3.40 | 3.12 | |
| DART_55023 | -14.80 | 153.58 | 2021.1-2021.12 | 3.45 | 4.64 | 3.39 | |
| Mean | | | | 3.88 | 4.19 | 3.97 | 0.56 |

Note: $M_2$ tide elevation RMS of EOT11a is from Stammer et al. (2014).

**Table 4. Hydrodynamic tide models' elevations against bottom pressure data in the Pacific Ocean**

|  | $M_2$ elevation RMS (cm) | Estimated total tide elevation RMS (cm) |
|---|---|---|
| HIM | 8.75 | 59.93 |
| OTIS-GN | 7.54 | 51.64 |
| STORMTIDE | 8.33 | 57.05 |
| OTIS-ERB | 5.63 | 38.56 |
| STM-1B | 12.69 | 86.92 |
| HYCOM | 7.82 | 53.56 |
| Mean | 8.46 | 57.95 |

Note: $M_2$ elevation RMS is from Stammer et al. (2014).

3.3.2 Tide-gauge data

We have put the model on the Pacific Ocean region, and thereby the island tide-gauge data may be another candidate to test it. The tide-gauge stations have been extensively distributed over this region, and the formed data are good quality and publicly accessible at some archives such as UHSLC (University of Hawaii Sea Level Center) (https://uhslc.soest.hawaii.edu/). At this time, we want a comparison of the prediction made by this model and the prediction made by harmonic analysis, which is already widely adopted in the tide field. Nevertheless, most of these archives provide only water level observation. In contrast, the NOAA (National Oceanic And Atmospheric Administration) of U.S.A is an ideal source from where both water level observation and water level prediction are provided. In the network of NOAA, 9 tide-gauge stations are presently installed in the Pacific Ocean region. Since the locations of 5 stations such as Nawiliwili, Honolulu, Mokuoloe, Kahului, and Hilo are close to each other, Honolulu station is finally considered. As a result, a total of 5 stations are selected for this study. A geographical distribution of these five stations is exhibited in Figure 8. First, we use hourly water level data of two months and hourly vertical displacement data of same duration to resolve the parameters of equation (4). The vertical displacement is also computed through equation (1). Then, we use equation (4) of known parameters and vertical displacement data to reproduce the water level prediction in the past time, and compare them to the observed water level and to the water level prediction made by the harmonic analysis. The time series of observation used to resolve equation (4) are from April 1 to May 31 of 2021, while the prediction series are from January 1 to June 30 of 2008. Table 5 exhibits a comparison of two predictions out to observation, with a further comparison shown in Figure 10. Over these 5 stations, their averaged *RMS* of prediction against observation for our model is 10.29 cm, while that for the harmonic analysis is 12.0 cm. The results of two sets of predictions are close, this proves our model to be good as the harmonic analysis. Please note, our model proceeds a backward prediction, and its duration reaches 13 years. This proves again our model to be constantly effective.

**Table 5. Tide elevation RMS comparison of OBOTM and Harmonic analysis in the Pacific Ocean**

| Island tide-gauge station | | | Observation series for resolving the parameters of equation (4) | Prediction series | RMS (cm) | |
|---|---|---|---|---|---|---|
| ID Station | Latitude (º) | Longitude (º) | | | OBOTM | Harmonic analysis |
| 1612340 Honolulu | 21.31 | 202.13 | 2021.4-2021.5 | 2008.1-2008.6 | 4.64 | 4.19 |
| 1619910 Sand Island | 28.21 | 182.64 | 2021.4-2021.5 | 2008.1-2008.6 | 7.34 | 9.85 |
| 1630000 Apra Harbor | 13.44 | 144.66 | 2021.4-2021.5 | 2008.1-2008.6 | 16.39 | 22.02 |
| 1820000 KWAJALEIN ATOLL | 8.73 | 167.74 | 2021.4-2021.5 | 2008.1-2008.6 | 14.64 | 8.89 |
| 1890000 Wake Island | 19.29 | 166.62 | 2021.4-2021.5 | 2008.1-2008.6 | 8.42 | 15.05 |
| | | | | | 10.29 | 12.00 |

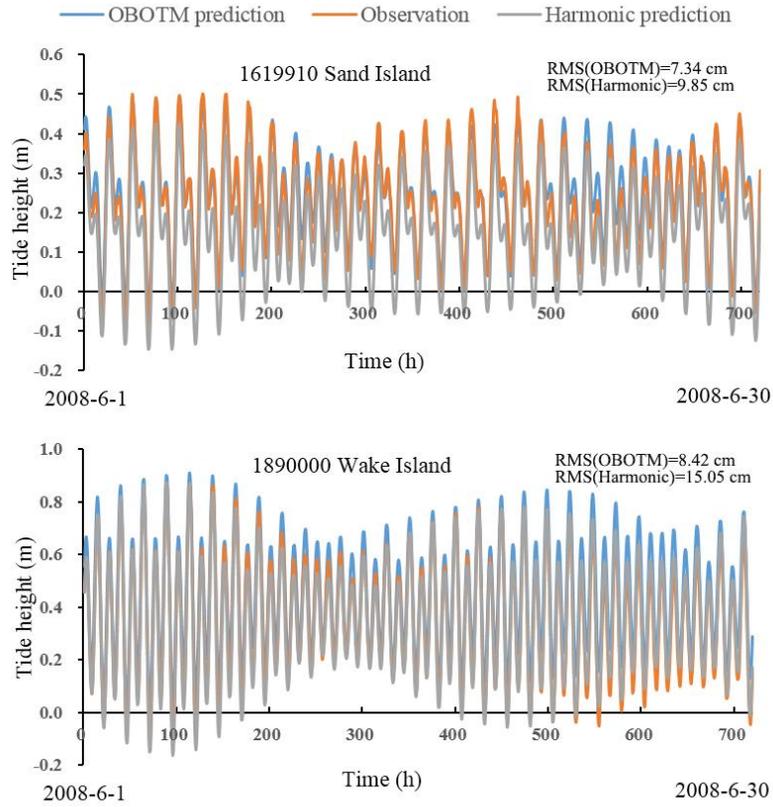

**Figure 10. Tidal elevation prediction comparison of OBOTM and harmonic analysis out to observation.** Tide-gauge data are from NOAA, where harmonic predictions are accordingly provided. The data points are hourly.

## 4. Oceanic basin oscillation-driving mechanism versus gravitational forcing mechanism

In section 3.3, we have shown quantitatively that the oceanic basin oscillation-driving mechanism is more consistent with the observed tides than the gravitational forcing mechanism. Even so, a qualitative argument on these two mechanisms is still necessary. As shown in Figure 11 (a1), the gravitational forcing mechanism begins with a derivation of the tide-generating force through a combination of the gravitational force and the centrifugal force, and this force is further decomposed into a horizontal force (called tractive force) and a vertical force. As the tractive force is exerted along the Earth's surface, applying this force to the whole ocean (Fig. 11(a2)), the observed tides are then thought to be a manifestation of the response of ocean to the force. Under the frame of the tractive force, we can expect that, along the longitudinal direction, the water at high latitude would be dragged to move towards low latitude, and along the latitudinal direction, all the water would be dragged to move towards west and east twice per day as the two tractive force vector fields are spatially symmetric. Apparently, the water movements $v_1$, $v_2$, $v_3$, $v_4$, $v_5$, $v_6$, $v_7$, and $v_8$ (Fig. 11(a3)) may be seen as representatives of some of the water movements initiated by the tractive force. In the northern ocean basin, the water movements $v_1$, $v_2$, and $v_3$ travel towards southeast, south, and southwest, respectively. However, they would be deflected by the Coriolis force and/or reflected by the continent, causing them to shift to $v_1$', $v_2$', and $v_3$', which finally travel towards southwest, southwest, and southeast, respectively. Similarly, in the southern ocean basin, the water movements $v_4$, $v_5$, and $v_6$ travel towards northwest, north, and northeast, respectively. After being deflected by the Coriolis force and/or reflected by the continent, they change direction to $v_4$', $v_5$', and $v_6$', which finally travel towards northeast, northwest, and northwest, respectively. Because of the obstacle of landmass, the larger tides would be expected to occur at the western and eastern boundaries of the ocean basin that is at low latitude, while the smaller tides would be expected to occur at the northern and southern boundaries of ocean basin. In contrast, as shown in Figure 11 (b1), the oceanic basin oscillation-driving mechanism treats the Earth as a whole, by applying the gravitational force and the centrifugal force to the Earth's body, solid Earth is mechanically elongated along the Earth-Moon line. As the Earth spins around its axis, the double solid bugle shakes ocean basin regularly (Fig. 11(b2)), forming water movements and water level changes around the globe (i.e., tides). Under the frame of the spinning double bulge, we may expect that, along the longitudinal direction, the water at low latitude would be forced to move towards high latitude; with the drop of the previously raised ocean floor, the water is induced to return from high latitude to low latitude, and along the latitudinal direction, all the water would be forced to move from east to west as ocean floor is orderly raised by the double bulge from east to west; with the drop of the previously raised ocean floor, the water is induced to return from west to east. Apparently, the water movements $v_1$, $v_2$, $v_3$, $v_4$, $v_5$, $v_6$, $v_7$, and $v_8$ (Fig. 11(b3)) may be seen as representatives of some of the water movements initiated by the double bulge. In the northern ocean basin, the water movements $v_1$, $v_2$, and $v_3$ travel towards southwest, south, and southwest, respectively. However, they would be deflected by the Coriolis force and/or reflected by the continent, causing them to shift to $v_1'$, $v_2'$, and $v_3'$, which finally travel towards southwest, southeast, and southeast, respectively. Similarly, in the southern ocean basin, the water movements $v_4$, $v_5$, and $v_6$ travel towards southwest, south, and southwest, respectively. After being deflected by the Coriolis force and/or reflected by the continent, they change direction to $v_4'$, $v_5'$, and $v_6'$, which finally travel towards southeast, southeast, and southwest, respectively. Because of the obstacle of landmass, the larger tides would occur at the any boundary of ocean basin, while the smaller tides would be expected to occur in the ocean.

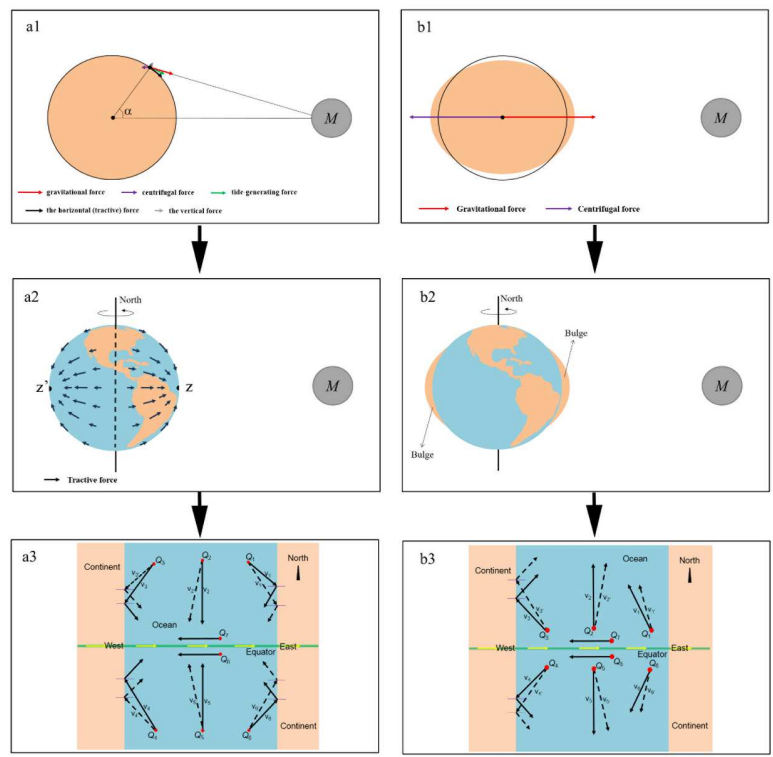

**Figure 11. Conceptual model of gravitational forcing mechanism (left) and oceanic**

**basin oscillation-driving mechanism (right).** a1, the derivation of the tide-generating force, α is the angle between the site and the Moon relative to the Earth's center (black dot). a2, the tractive force around the globe, Z and Z' are locations of sub-lunar points at the Earth's surface. a3, the water movements ($v_1$, $v_2$, $v_3$, $v_4$, $v_5$, $v_6$) initiated by the tractive force and the water movements ($v_1'$, $v_2'$, $v_3'$, $v_4'$, $v_5'$, $v_6'$) further influenced by continent and the Coriolis force. b1, the elongated solid Earth due to a combination of the gravitational force of the Moon and the centrifugal force, black circle denotes the original shape of solid Earth. b2, the double solid bulge around the globe. b3, the water movements ($v_1$, $v_2$, $v_3$, $v_4$, $v_5$, $v_6$) initiated by the double bulge and the water movements ($v_1'$, $v_2'$, $v_3'$, $v_4'$, $v_5'$, $v_6'$) further influenced by continent and the Coriolis force. $Q_1$, $Q_2$, …, and $Q_8$ in a3(b3) denote the initial positions of water movements, yellow line with arrow in a3(b3) denotes that the Earth spins from west to east.

Now, let's take these expectations to see which of two mechanisms is apparently more suitable for the observed tides around the globe. As shown in Figure 12, the larger tides occur generally at coasts, while the smaller tides occur mostly in oceans. At most of coasts (see Figure 12(a)), the tidal ranges are generally more than 2.0 m, while in most of oceans, they are less than 1.0 m. In Alaska, North Sea, the northern boundary of the Indian Ocean, and the northwest boundary of the Pacific Ocean, the tidal ranges reach up to 4.0~6.0 m. The world largest tide occurs in the Bay of Fundy, where the tidal range are greater than 11.0 m. At the port of Anchorage of Alaska and at the head of California gulf, the tidal ranges are close to 11.0 and 7.0 m, respectively (Segar, 2018). The Bay of Fundy is located on the east coastline of North America and is a funnel-shaped inlet of the Atlantic Ocean, its trendy is from southwest to northeast, while the California gulf is located on the west coastline of North America and is a large inlet of the eastern Pacific Ocean, its trendy is from southeast to northwest. The large tides occurred at these places would require water to move towards north (from low latitude to high latitude). In the oceans (see Figure 12(b)), the co-tidal lines appear mostly perpendicular to the coastlines, and several large amphidromic systems dominate most of oceans, with tidal phase performing counter-clockwise (clockwise) rotation in the northern (southern) hemisphere. These patterns also would require water to move from low latitude to high latitude, and with aid of the Coriolis force and/or the continent's reflection (refraction), the water movements are rotated. All these points are clearly contradictive with the expectations of the gravitational forcing mechanism. In contrast, the occurrences of the large tides at the northern boundary of ocean basin and the large amphidromic systems in the oceans are well consistent with the expectations of the oceanic basin oscillation-driving mechanism.

If we think the matter more deeply, some key problems may be found from the frame of the tractive force. Energetically, the tractive force represents a potential field that is uniformly directed towards the sub-lunar points (such as *m* and *m'*, which are projection of the Moon on the Earth's surface) (see Figure 11($b_2$)). Within this field, the tractive force pulls water to move along the force vector. As the sub-lunar points are mostly located at low latitudes (typically between 18º N and 18º S), and the Earth spins around its axis, all the water would be dragged to move from east to west and from west to east twice per day, while those at middle and high latitudes would be pulled to move towards the low latitude. Additionally, the sub-solar points (i.e., projection of the Sun on the Earth's surface) are also located at low latitudes (typically between 23.5º N and 23.5º S), and the tractive force generated by the Sun is largely less than that generated by the Moon, as a result, the inclusion of the tractive force generated by the Sun does not change the pattern dominated by the tractive force generated by the Moon. However, as described in these standard oceanographic textbooks (i.e., Pugh and Woodworth (2014); Gerkema (2019)), the tidal waves are initiated in the deep oceans from where they entrain energy to propagate towards global coasts. For instance, in the northwest American shelf seas, the tidal wave travels northward along the coast from south to north. In the Yellow Sea, the tidal wave enters from the East China Sea, with the largest amplitude at the coast of Korea, and the returning wave travels southward along the coast of China. In the northwest European shelf, tides approach from the Atlantic Ocean to the north, taking about 5 hours to travel from the Celtic Sea to the Shetlands, and to the east, taking about 7 hours to travel from the shelf edge to the head of the Irish Sea. In the Gulf of Maine and the Bay of Fundy, tidal amplitude increases from less than 0.5 meters at the shelf edge to over 5.64 meters at Burncoat Head. Further north, the tidal wave enters the Gulf of St Lawrence via the Cabot Strait (Forrester, 1983), causing the tidal range to increase as it moves up the St Lawrence River. We must recognize, these northward tidal progressions essentially represent some energy flows whose directions are opposite to the directions of the tractive force. In addition to this, as the tractive force always pulls water to move towards low latitude, there would be an accumulation of water at low latitude and a shortage of water at high latitude. This would finally result in an imbalance of water between the ocean at high latitude and the ocean at low latitude. However, the ocean water is practically balanced everywhere. In contrast, the double bulge may be easy to account for these observed tidal dynamics. For instance, as shown in Figure 11(b2), the double bulge is mostly located at low latitude, and the Earth spins around its axis, when the middle part of the Pacific (Atlantic ) Ocean floor is raised by the solid bulge, the raised ocean floor would cause the water above it to radically flow out, consequently, some of water would travel towards north, west, and south. When the travelling water (i.e., tidal wave) meets landmass, a reflection or refraction takes place, this finally changes its path. Since the ocean floor is consecutively raised/ dropped by the two bulges, the dropped ocean floor would induce a return of water to compensate the water lose, this process guarantees a balance of water in the ocean.

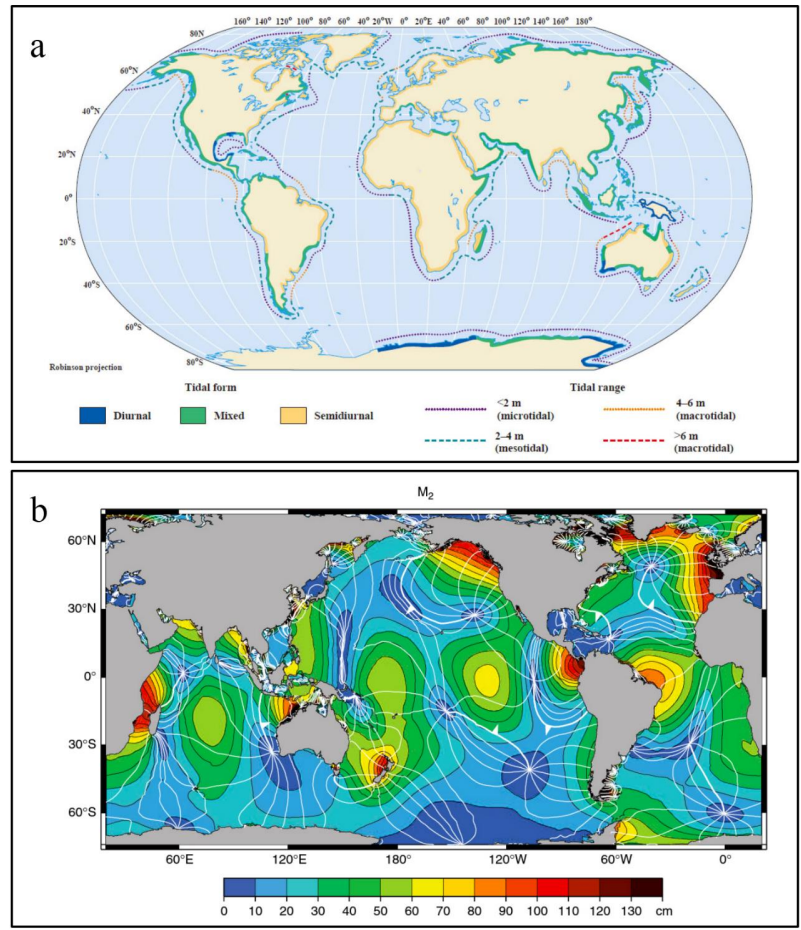

**Figure 12. Global tide distribution.** a, tides along coasts. "Douglas A. Segar, Introduction to Ocean Science, 2018; licensed under a Creative Commons Attribution (CC BY) license." b, $M_2$ co-tidal charts of the ocean tide. "Reproduced with permission from Richard Ray/Space Geodesy branch, NASA/GSFC."

## 5. Discussion

It is widely believed that the existing tidal theory provides accurate tide predictions through harmonic analysis, which in turn proves the existing tidal theory to be correct. This view deserves a wider discussion. On the one hand, there is no standard for tide prediction accuracy. As described in the documented file of NOAA, observed tides are always a combination of astronomical tides and non-tidal variances (such as weather and wind), which are irregular and unpredictable; Therefore, tide predictions cannot account for these variances. Comparing tide predictions to observations can result in a wide range of discrepancies, from a few centimeters to tens of centimeters or more in accuracy. More detail of this variance may refer to the published tide table (https://tidesandcurrents.noaa.gov/historic_tide_tables.html). On the other hand, the harmonic prediction is based on mathematics rather than physics. As introduced in Fok (2012), the harmonic prediction begins with an expansion of the tidal potential that is a function of the time-dependent position of the attracting bodies (i.e., Moon, Sun). The motion of any of these bodies can be described by six astronomical angles approximately linear with time. Incorporating these parameters into the potential equation, the effects of these parameters on a location are expressed by a set of corresponding harmonics (Smith, 1999). Further, each of the harmonics is written into a sinusoid of two unknown constants (i.e., amplitude and phase), which are subsequently resolved through a least-square fitting to the observed tides at this location. Last, by calculating the variation of time in these sinusoids of known constants, a simple addition of these constituents works out the predicted tides at this location in the future (Gerkema, 2019). The harmonic prediction at a location is conventionally expressed as below (see Hibbert et al. (2015)):

$$T(t) = Z_0 + \sum_{n=1}^{N} H_n f_n \cos[\sigma_n t - g_n + (V_n + u_n)] \qquad (7)$$

Where T($t$) denotes the observed tidal elevation at times ($t$), $Z_0$ is the mean water level, $H$ and $g$ are tidal amplitude and phase lag, respectively, $\sigma$ and $V_n$ are angular speed and equilibrium phase angle of the $n$th tidal constituent, respectively, and $f_n$ and $u_n$ are nodal factors.

We find that the harmonic prediction is firstly applying each of these parameters (i.e., external forces) to the given location to produce a water level change, and then using a sum of all water level changes produced by these parameters to represent the final water level change at the location. Clearly, this method has separated the location from the whole ocean. Ocean water is a continuous mass, and is also incompressible. This situation means that the water at a location would not simply respond to any external force that applies to it, and the final water level change at the location is a consequence of the water level changes at all the locations of the whole ocean. But why may the harmonic prediction be successful? It is simple because any oscillation (tide, for instance) may be mathematically expressed with a sinusoid of time, and a sinusoid can be further decomposed into many sinusoids of time. The harmonic prediction skillfully makes use of this feature of tide, so, it's destined to succeed. We was frequently asked why don't use the newly presented model OBOTM to compare with the currently best tide modes like EOT20 and Fes2014. The answer to this question is also simple. First of all, these models are built based on observations through harmonic analysis that is rather mathematical (i.e., without a foundation of physics). Second, we here present a new tidal theory and develop a tide model OBOTM based on this theory, if a quantitative comparison of this new tidal theory and the existing tidal theory is needed, we should compare OBOTM with the hydrodynamic modes rather than these data-constrained models like EOT20 and Fes2014, since the hydrodynamic models are the right representative of the existing tidal theory.

As shown in these purely hydrodynamic models (i.e., HIM, OTIS-GN, STORMTIDE, TiME), several primary items (i.e., the tidal forcing, the self-attraction and loading, and topographic drag) have been included. The tidal forcing results from the gravitational force but is modified by a factor of $1+k_2-h_2$, where the Love numbers $h_2$ and $k_2$ respectively account for earth tide and its perturbation gravitational potential (Arbic et al., 2004). Even so, as we revealed in Table 4, the tidal elevation accuracy of these models is rather low (their *RMS*s reach up to tens of centimeters, for instance). Contrary to these hydrodynamic models, our model OBOTM has realized a better elevation accuracy (see Table 2) by neglecting the primary items mentioned above. This reality indicates that a simplified treatment of the complex situation may help realize a better understanding of ocean tide.

The number of the primary sites that are used to produce vertical displacement (i.e., water depth change) determines the accuracy of our model. We have adapted a grid of 10º resolution, and the locations of node of this grid are treated as the primary sites. If this grid is further subdivided, more primary sites would be extracted to improve the model's accuracy, but this would require a strong computing power that cannot be endured at present. In this study, we have used the spinning double bulge to explain ocean tides in both qualitative and quantitative manners. The water movements (i.e., tidal waves) initiated in oceans would finally arrive at the shelf seas and coastlines, where the interaction of ocean tide and local oscillation is remarkable and shallow water dynamics plays an important role. Next, we will put our model on these regions to see its applicability.

The existing understanding of tides includes that the Moon's (Sun's) gravitational force impels ocean water to form ocean tide and deforms solid Earth (known as earth tide or body tide), ocean tide loading contributes to solid Earth deformation, earth tide also has influence on ocean tide, and that the contribution of earth tide to ocean tide at a location is a simple addition of equilibrium tide amplitude, which can expressed by $(1+k-h)\Omega p/g$, where $\Omega p/g$ is the equilibrium tide amplitude at the location and $(1+k-h)$ is a combined diminishing factor in the equilibrium tide (Pugh and Woodworth, 2014). Ocean water is a continuous mass, the occurrence of earth tide (solid Earth deformation) would lead ocean basin to shake regularly, the water level change at a location must be a consequence of the water level changes across all oceanic locations. Evidently, the existing understanding of tides fails to include the influence of oscillating ocean basin on ocean water. Undoubtedly, the ocean basin oscillating-driving mechanism presented in this study is filling this blank. The future aspects of this present work will encompass: 1) We have presented a new tidal theory (termed the oceanic basin oscillating-driving mechanism) and have developed a tide model OBOTM based on this theory. However, it is worth noting that the model in its current form remains relatively simplistic, as it excludes the influences of ocean depth, landmass, the Coriolis force, and friction on water movement. From this side, there is a need for the development of a more comprehensive hydrodynamic model grounded in this present model. 2) As demonstrated in section 3.2 of this work, ocean basin is being regularly shaken in the elongated spinning solid Earth. Yang et al. (2024) have unveiled that solid Earth deformation (earth tide) varies in a timely manner. As a part of solid Earth, ocean basin would be entrained to vary its volume accordingly. Furthermore, this variation would modify global water level. However, the impact of this varying ocean basin on water level is not yet considered in determining sea level changes. Hence, the current understanding of sea level change needs to be reevaluated.

## 6. Conclusion

Tides represent the daily alternations of high and low waters along coastlines and in oceans, explaining the physics behind them has challenged human wisdom over thousands of years. The current theory (termed the gravitational forcing mechanism) explains tides as a manifestation of the response of ocean water to the gravitational force of the Moon and Sun. Alternatively, in this study we present a new theory (termed the oceanic basin oscillation-driving mechanism), in which tides are explained as a manifestation of oscillating ocean basin that is intricately linked to the elongated spinning solid Earth. Based on this theory, we have developed an Oceanic Basin Oscillation Tide Model (OBOTM) and tested it using 11-year observations from 33 bottom pressure stations across the Pacific Ocean, the average Root Mean Square (*RMS*) deviation of tidal elevation (representing water depth change) predicted by this model, compared to observations, is 7.54 cm. Using a ratio of $M_2$ elevation *RMS* of ocean tide model EOT11a and its total tidal elevation *RMS* as a reference, we estimated the total tidal elevation *RMS* of six purely hydrodynamic models (HIM, OTIS-GN, STORMTIDE, OTIS-ERB, STM-1B, and HYCOM), which are representatives of the current theory, to be 59.93, 51.64, 57.05, 38.56, 86.92, and 53.56 cm, respectively. Furthermore, we compared OBOTM with harmonic prediction derived from observations at five tide-gauge stations. OBOTM demonstrated an average *RMS* deviation of tidal elevation of 10.29 cm against observations, while the harmonic prediction demonstrated an average *RMS* deviation of 12.00 cm. We also show that the new theory is apparently more suitable for explaining the observed tide dynamics in the Pacific and Atlantic Oceans than the current theory. Our findings suggest that the new theory offers a more compelling explanation for tides compared to the current theory, and will greatly contribute to many fields such as oceanography and geophysics.

**Supplementary Material**

This includes an excel code of the newly presented earth tide model (Yang et al., 2024), and the code allows readers to easily produce tidal displacement of any location during a period of 1990-2030.

**Acknowledgements**

We are very grateful to Philip Woodworth, John Huthnance, XianQing Lv for their helpful comments on this work. We thank Walter Babin, Thierry De Mees, Roger A. Rydin, and Wouter Schellart for their suggestions on earlier versions of the text. It is grateful to Richard Ray for providing the image of $M_2$ tide. We also thank Ole Baltazar Andersen and Torsten Mayer-Gürr for helping produce ocean tide data through models.

41